\documentclass[fleqn,usenatbib]{mnras}

\usepackage[T1]{fontenc}

\DeclareRobustCommand{\VAN}[3]{#2}
\let\VANthebibliography\thebibliography
\def\thebibliography{\DeclareRobustCommand{\VAN}[3]{##3}\VANthebibliography}

\usepackage{graphicx}	
\usepackage{amsmath}	
\usepackage{amssymb}	

\usepackage{soul}
\usepackage[normalem]{ulem}

\defcitealias{Garon2019}{G19}
\defcitealias{WenHan2015}{WH15}

\usepackage[dvipsnames]{xcolor}

\usepackage[nameinlink, capitalise, noabbrev]{cleveref}
\usepackage{newtxtext,newtxmath}

\title[Clusters' influence on NATs]{Clusters' far-reaching influence on narrow-angle tail radio galaxies}

\author[K. de Vos et al.]{
K. de Vos,$^{1}$\thanks{E-mail: ppykd1@nottingham.ac.uk (KdV)}
N. A. Hatch,$^{1}$
M. R. Merrifield,$^{1}$
B. Mingo,$^{2}$
\\
$^{1}$School of Physics and Astronomy, University Park, University of Nottingham, Nottingham, NG7 2RD, UK\\
$^{2}$School of Physical Sciences, The Open University, Walton Hall, Milton Keynes, MK7 6AA, UK}

\date{Accepted XXX. Received YYY; in original form ZZZ}

\pubyear{2021}

\begin{document}
\label{firstpage}
\pagerange{\pageref{firstpage}--\pageref{lastpage}}
\maketitle

\begin{abstract}

In order to study the ram-pressure interaction between radio galaxies and the intracluster medium, we analyse a sample of 208 highly-bent narrow-angle tail radio sources (NATs) in clusters, detected by the LOFAR Two-metre Sky Survey. For NATs within $7\,R_{500}$ of the cluster centre, we find that their tails are distributed anisotropically with a strong tendency to be bent radially away from the cluster, which suggests that they are predominantly on radially inbound orbits. Within $0.5\,R_{500}$, we also observe an excess of NATs with their jets bent towards the cluster core, indicating that these outbound sources fade away soon after passing pericentre. For the subset of NATs with spectroscopic redshifts, we find the radial bias in the jet angles exists even out to $10\,R_{500}$, far beyond the virial radius. The presence of NATs at such large radii implies that significant deceleration of the accompanying inflowing intergalactic medium must be occurring there to create the ram pressure that bends the jets, and potentially even triggers the radio source. 

\end{abstract}




\section{Introduction}
\label{sec:Introduction}

The interaction between galaxies and their surrounding gas, whether circumgalactic medium, intergalactic medium or intracluster medium (ICM), is a major driver of galaxy evolution. Nowhere is this interaction more dramatically demonstrated than in radio galaxies moving through the ICM of their surrounding cluster. The synchrotron-emitting plasma that comprises the radio lobes is ejected from the body of the galaxy, meaning that it is subject to the hydrodynamical processes that result from its interaction with the ICM: ram pressure will cause the jets to bend \citep{Cowie1975, Begelman1979, O'Dea1985, Roberts2021}, while buoyancy effects can cause them to "float" towards the edge of the cluster \citep{Gull1973, Gendron-Marsolais2017}.

Narrow-angle tail radio sources (NATs) are a particular class of extended, double-tail radio galaxy that have had their radio jets bent back such that the observed angle between them is acute. Of course, this projected angle on the sky may not reflect the true three-dimensional bending which could be significantly less extreme, but the most plausible underlying physical cause of such distortions -- ram pressure due to the galaxy's motion relative to the cluster -- means that this projected geometry can always be used as a diagnostic of the source's direction of motion on the plane of the sky. \citet{O'Dea1987} took this approach to analyse the orbits of 70 NATs in Abell clusters, and concluded that the orbits were close to isotropic, but with some indication of a radial bias at small radii.  However, they argued that a larger sample was required to make any definitive statement about cluster orbits.  

The largest study of bent radio jets in clusters to date is by \citet[hereafter \citetalias{Garon2019}]{Garon2019}, who made use of a sample of extended radio galaxies identified through Radio Galaxy Zoo \citep{Banfield2015}. \citetalias{Garon2019} found that the 340 radio sources they identified as "highly bent" have a slight tendency to indicate radial orbits with respect to their cluster centre. They also discovered that such bent systems were found out to fairly large radii, with as many outside $1.5R_{500}$ as inside it. Since ram pressure is proportional to the density of the ICM, and is a necessity for bending double-tail radio sources to such a high degree, it is puzzling that such bent sources would be commonly found out at large distances from the cluster centre where the ICM density is low.

However, \citetalias{Garon2019} adopted a generous limit on what constituted a "highly bent" double-tail source, and in fact explicitly excluded all sources in which the observed angle between the two radio jets was less than $45^\circ$ over the concern that such objects might be mis-associated background sources. Since these steeply-bent sources comprise a large proportion of what is classically labelled as a NAT, it is not clear that \citetalias{Garon2019} and \citet{O'Dea1987} identified comparable populations, and hence whether the physics bending the jets is the same in both cases.

We have therefore sought to revisit these issues, focusing specifically on radio galaxies identified as NATs in the largest sample available to date. We explore the angles in which their jets are bent relative to the closest cluster, and investigate in more detail how this distribution varies with projected distance from the cluster centre. In \cref{sec:Data_Method}, we describe the data set and analysis technique, while \cref{sec:Results} presents the resulting distribution of orbital angles and its variation with radius, and compares it to the analysis of \citetalias{Garon2019}. In \cref{sec:Discussion}, we discuss the implications of the rather unexpected but very strong signal that we detect.

\section{Data and Method}
\label{sec:Data_Method}

For this study, we use images from the Low-Frequency Array (LOFAR) Two-metre Sky Survey (LoTSS) first data release (DR1; \citealt{Shimwell2019}). LoTSS DR1 is the ideal data set to identify NATs, as the 424 square degree high-resolution radio survey is not only an order of magnitude deeper than previous wide-area radio surveys, with a median noise level of only $71\mu{\rm Jy \,beam^{-1}}$, but is sensitive to structures with sizes ranging from 6\,arcsec to more than a degree. In addition, its observations at 144MHz have been shown to be significantly better for the detection of NATs than higher frequency data, due to the steep spectra of such mature radio sources \citep{O'Neill2019a, O'Neill2019b}.
 
From this data set, we extract the 264 NATs visually identified by \citet{Mingo2019}, who classified the morphologies of 5805 extended radio-loud AGN in LoTSS DR1. Optical or infrared counterparts for all of the NATs were identified by \citet{Williams2019} using either a likelihood ratio identification algorithm or, for larger and more complex sources, visual identification through the LOFAR Galaxy Zoo project\footnote{\url{https://www.zooniverse.org/projects/chrismrp/radio-galaxy-zoo-lofar}}. Spectroscopic redshifts are available for 179 of the 264 NATs, and for the remaining 85 NATs we use the photometric redshifts derived by \citet{Duncan2019}, which have an overall scatter $\sigma_{\rm NMAD}=0.039$ and an outlier fraction of 7.9\%.

To identify the environments of the NATs, we draw on the cluster catalogue by \citet{WenHan2015}, selected from the optical SDSS DR12 catalogue \citep{Alam2015}. This catalogue contains 158,103 clusters in the redshift range $0.02<z<0.8$, identified using a friends-of-friends (FoF) algorithm. The catalogue is 95\% complete for clusters of mass $M_{200} > 10^{14} M_\odot$ (where $M_n$ is the mass inside radius $R_n$, within which the density is $n$ times the critical density of the Universe), and it has a false detection rate of less than 6\%.
The centre of each cluster is defined by the position of the brightest cluster galaxy (BCG), which is identified as the brightest galaxy within 0.5\,Mpc and a redshift within $\pm 0.04(1 + z)$ of the densest region of each cluster identified by the FoF algorithm.

92.6\% of the clusters within the 424 square degree region covered by LoTSS DR1 have spectroscopic redshifts, which are defined as either the redshift of the BCG or the mean redshift of the cluster members. The remainder of the clusters have photometric redshifts derived by \citet{WenHanLiu2012}, which have a standard deviation of less than $0.018$.  \citet{WenHan2015} determine $R_{500}$ and $M_{500}$ for each cluster using empirically-derived scaling relations, which we then use to obtain a characteristic velocity dispersion of each cluster, $\sigma_{500} \equiv (GM_{500}/R_{500})^{1/2}$. 

Following \citetalias{Garon2019}, we determine the most likely host cluster (if one exists) for a NAT at redshift $z_{\rm NAT}$ by first identifying all clusters whose redshifts $z_{\rm cluster}$ satisfy $|z_{\rm NAT}-z_{\rm cluster}|/(1+z_{\rm NAT})<0.04$, which, reflecting the photometric redshift uncertainties, corresponds to a velocity window of $\pm 12,000\,{\rm km/s}$. We then assign the host to be the cluster with the minimum projected distance to the NAT. These criteria associated a cluster with 255 of the NATS in our sample. In 47 cases, the optical source associated with the NAT was found to be the BCG of the cluster; since these objects are used as a proxy to define the centre of the cluster, there is no meaningful information to be obtained from them regarding offsets from the cluster centre. To exclude such objects, while allowing for possible centring errors, we eliminate all sources within a projected radius of $0.01R_{500}$, which leaves a final sample of 208 NATs.

\begin{figure}
	\includegraphics[width=\columnwidth]{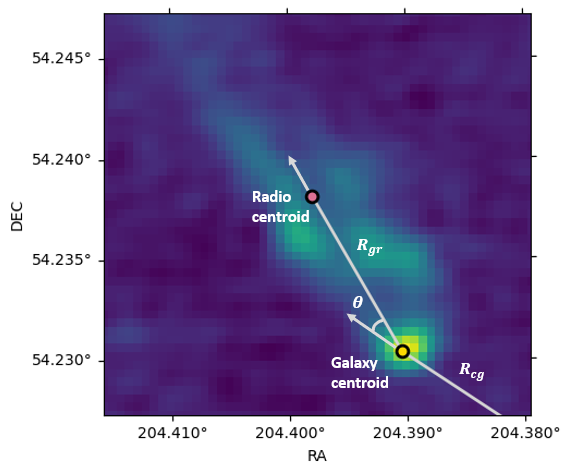}
    \caption{The definition of the angle $\theta$ between ${\mathbf{R}}_{cg}$ and $\mathbf{R}_{gr}$, overlaid on the image of a typical NAT as observed by LOFAR.}
    \label{fig:vector_angle}
\end{figure}

For each NAT--cluster pair, we calculate $\theta$, the counter-clockwise angle between the vector from the cluster centre to the galaxy centroid, ${\mathbf{R}}_{cg}$, and the vector from the galaxy centroid to the centroid of radio emission, $\mathbf{R}_{gr}$; a typical example of this calculation is shown in \cref{fig:vector_angle}, which also illustrates the high quality of the LoTSS data. We map these angles into the range $-180^\circ < \theta < 180^\circ$, so that $|\theta| \sim 0^\circ$ describes a radio tail pointed away from the cluster centre, while $|\theta| \sim 180^\circ$ represents a radio tail aligned toward the cluster centre. 

\section{Results}
\label{sec:Results}

\begin{figure}
    \includegraphics[width=\columnwidth]{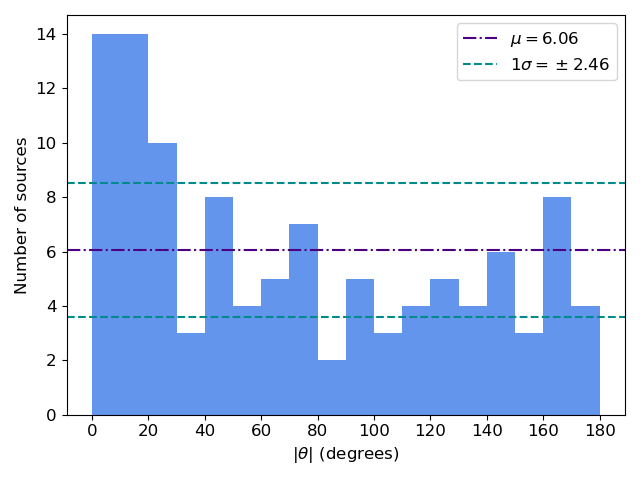}
    \caption{The angle distribution of narrow-angle tail radio sources with respect to their cluster centres, out to $7R_{500}$. The lines indicate the expectations of a uniform distribution, with Poisson noise appropriate to the size of the sample.}
    \label{fig:NAT_hist}
\end{figure}

Using a conservative limit of $R < 7 R_{500}$ to avoid significant line-of-sight contamination, we are left with a sample of 109 NATs, the angle distribution for which are presented in \cref{fig:NAT_hist}. It is immediately apparent from this figure that the data does not appear consistent with the expectations of a uniform distribution, but rather shows an excess at small angles. To test the significance of this apparent non-uniformity, we conducted an Anderson--Darling (AD) test, which offers a more powerful tool than the commonly used Kolmogorov--Smirnov test when assessing the significance of features near the ends of a distribution \citep{Stephens1974}. Testing the observed distribution against a uniform model results in an AD statistic of 7.94, which is significant at the 99.99\% confidence level for a sample of this size \citep{Jantschi2018}. To check that this result is not an artefact produced by the flux-weighted manner in which we have defined the NATs' angles on the sky, we repeated the analysis using the angle defined by the bisector of the two jets in each NAT, as located by their peak fluxes \citep{Mingo2019}; this definition produced a very similar non-uniform distribution of angles. We also note that any residual uncertainty in this measurement would serve only to dilute the signal apparent in \cref{fig:NAT_hist}.

\begin{figure}
     \includegraphics[width=\columnwidth]{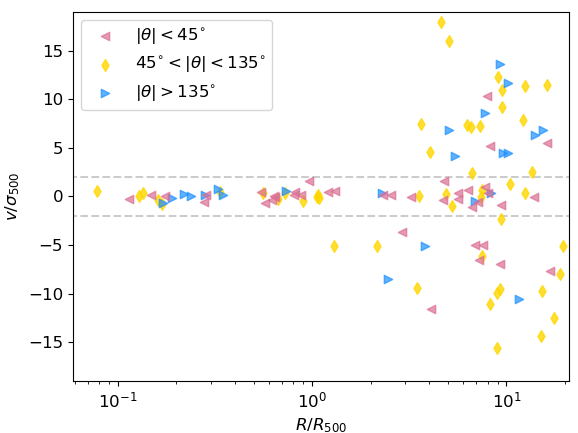}
    \caption{The projected phase space distribution (showing line-of-sight velocity versus projected separation) of the NATs in the sample with spectroscopic redshifts out to $20R_{500}$. The dashed lines at $v = \pm2\sigma_{500}$ indicate the limits of velocity assumed to be associated with the cluster.}
    \label{fig:phase_space}
\end{figure}

We next assess the level of line-of-sight contamination caused by using the photometric redshifts for the subsample of the NATs that lack spectroscopic data. \cref{fig:phase_space} shows the projected phase-space diagram for the subset of objects for which we have full spectroscopic redshifts. The data points have been scaled by their individual values of $R_{500}$ and the characteristic velocity $\sigma_{500}$, so that objects in clusters of differing mass can be compared consistently in this phase space.  Although the amount of line-of-sight contamination clearly increases with radius, this plot confirms that its level remains modest out to the $7 R_{500}$ limit adopted in \cref{fig:NAT_hist}: only $\sim 25\%$ of the cluster-NAT pairs are false associations which act to dilute the signal.

Beyond this radius, the level of contamination increases rapidly, but for these sources with spectroscopic redshifts we can extract a largely uncontaminated sample out to significantly larger radii by taking the NATs that lie within the dashed lines on \cref{fig:phase_space}, for which $|v| < 2\sigma_{500}$.  Reassuringly, as is also apparent from \cref{fig:phase_space}, the contaminating sources excluded by this process have an angle distribution that is consistent with random, confirming that the alignment effect in \cref{fig:NAT_hist} is associated with the cluster rather than some spurious systematic bias. 

In the remaining sources that are associated with clusters, the alignment effect appears to persist out to at least $\sim 10 R_{500}$. We confirm that this phenomenon is not just associated with the cluster core by repeating the AD test on the spectroscopically-confirmed associations that lie in the radial range $3R_{500} < R < 10 R_{500}$, well outside the virial radius, which lies at $\sim 1.4R_{500}$ \citep{Walker2019}. We find that their angular distribution is also inconsistent with a uniform distribution at the 99.9\% confidence level, with an AD statistic of 6.02.  Interestingly, if we do look at just the cluster core in \cref{fig:phase_space}, there also appears to be an excess of NATs for which $|\theta| > 135^\circ$, which we will discuss further in \cref{sec:Discussion}. 

It is notable that these results differ from those of \citetalias{Garon2019} in several ways. Not only is the non-uniformity in $\theta$ presented in \cref{fig:NAT_hist} significantly stronger that that detected by \citetalias{Garon2019}, but it also shows up as an asymmetric feature; we find that many more tails are directed away from the cluster than toward it, whereas \citetalias{Garon2019} determined the presence of this asymmetry in folded data but did not disaggregate these populations. In addition, we find strong evidence that this phenomenon persists to much larger radii than previously probed. The greater strength of these effects suggests that the LOFAR-detected NATs studied here are more dramatic probes of the ICM--galaxy interaction than previously recognised.

\section{Discussion}
\label{sec:Discussion}
\begin{figure}
    \includegraphics[width=\columnwidth]{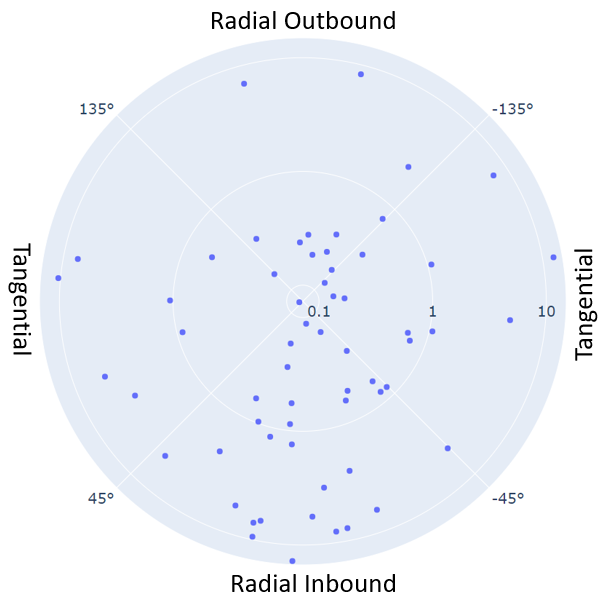}
    \caption{A polar diagram of the distribution of NAT angles on-the-sky, $\theta$, as a function of radius, $R/R_{500}$, for those sources spectroscopically confirmed to be associated with a cluster, such that $|v| < 2\sigma_{500}$. The cardinal points are labelled to show the orbital direction we would expect the galaxies to be travelling in with respect to the cluster centre, if the values of $\theta$ are the result of ram-pressure bending of their jets.}
    \label{fig:polar_plot}
\end{figure}

In this analysis, we have shown that the angle distribution of NATs implies that at least some are aware of the direction to the nearest cluster of galaxies, and that this awareness extends to surprisingly large distances.  By way of summary, \cref{fig:polar_plot} shows in polar form how, for the spectroscopically-confirmed NATs, the angles between bent radio jets and cluster centres are distributed as a function of radius. This plot emphasizes the preference of the radio jets to point away from the cluster out to $\sim 10 R_{500}$, but also indicates the secondary feature of an excess of NATs whose tails point toward the cluster centre within $\sim 0.5 R_{500}$. The cardinal labels on \cref{fig:polar_plot} indicate the direction of travel of the NATs on the plane of the sky, assuming that their tail-bending is due to ram pressure.

Such features are notable because even if an infalling radio galaxy and its immediate surroundings are close enough to feel the gravitational effects of the nearby cluster, the equivalence principle implies that they cannot be aware of such influence -- and hence the jets cannot be bent in specific directions -- if they are simply freely falling in that gravitational field. As previous studies of jet bending have noted, it requires the additional presence of hydrodynamical phenomena, where forces other than gravity are in play \citep{Cowie1975,Begelman1979, O'Dea1987, Sakelliou2000}.

Within the virial radius of a cluster, one would expect the ICM to be largely in hydrostatic equilibrium, so radio jets emerging from galaxies in this region would be bent by their motions relative to this stationary gas due to ram pressure. Any additional infalling gas is rapidly decelerated at a "virial shock" close to this radius \citep{Hurier2019}, although the morphology of shocks in infalling gas can be quite complex, with external shocks occurring all the way out to $\sim 5 R_{500}$ \citep{Molnar2009, Walker2019}. Such shocks produce the non-gravitational changes in the bulk flow of the gas that decouples it from the motions of galaxies, potentially providing the speed differential required to form NATs.

However, none of these shock processes seem to be predicted to occur out to the $\sim 10 R_{500}$ where NATs are observed here.  We therefore suggest that the true morphology of infalling gas is yet more complex, with significant hydrodynamical processes occurring out to even larger radii.  In this context, it is interesting to note that Fig.~1 of \citet{Reiprich2013} shows tendrils of heated gas extending to well beyond $5 R_{500}$.  While such radially-extended features may be quite rare, it seems likely that hydrodynamical phenomena also play a role in triggering the AGN activity in the first place \citep{Poggianti2017, Marshall2018, Ricarte2020}, which means that radio jets will preferentially be generated in just these regions, highlighting where they do occur.

An infalling NAT will, after passing its pericentre near the cluster centre, then continue radially outward on its orbit, on its way to becoming a cluster member.  Indeed, we can seemingly identify such a component in \cref{fig:polar_plot}, which shows an excess of NATs within $R_{500}$ that have their jets bent at $|\theta| \sim 180^\circ$, indicating an outbound galaxy on the plane of the sky if the bends are caused by ram pressure. It is interesting to note that the timescale on which an outbound galaxy will reach the $\sim0.5R_{500}$ radius at which the NATs seem to fade out, $\tau \sim 0.5R_{500}/\sigma_{500}$, is, for the characteristic masses of clusters in this sample, a few hundred million years, which is directly comparable to the lifetimes predicted for such sources \citep{Antognini2012}. This coincidence suggests that pericentre passage may represent the point at which new NATs are no longer being triggered.

We thus have a scenario that at least plausibly explains the rather unexpected structures apparent in \cref{fig:polar_plot}. At large radii, galaxies and gas lie in infalling filaments, some of which are dense enough that hydrodynamic effects start to decelerate the gas relative to the galaxies. The resulting differential will disturb the gaseous environment around the galaxies, potentially triggering AGN activity to produce large-scale radio jets, and these jets are then bent through ram pressure effects that arise from the speed differential. At least some of these radio jets have long enough lifetimes to survive their pericentre passage, creating the excess of radially outbound NATs at small radii. It would be very interesting to see this picture fleshed out in further detail both by adding more data from the ongoing LOFAR surveys, and from a full comparison to simulations of cluster evolution that incorporates both detailed gas physics and the triggering of AGN activity.  

\section*{Acknowledgements}
We thank G.\,K.~Miley, Huub R\"ottgering, D.\,J.\,B.~Smith and T.\,W.~Shimwell for their helpful comments on this work. 
KdV, NAH and MRM acknowledge support from the UK Science and Technology Facilities Council (STFC) under grant ST/T000171/1, and BM acknowledges support from the UK STFC under grants ST/R00109X/1, ST/R000794/1, and ST/T000295/1.
All authors are grateful for the use of data from LOFAR, the LOw Frequency ARray, and SDSS, the Sloan Digital Sky Survey.
This research made use of Astropy, a community-developed core Python package for astronomy \citep{astropy:2013, astropy:2018} hosted at \url{http://www.astropy.org/}, of MATPLOTLIB \citep{Hunter:2007}, of Plotly \citep{plotly}, and of TOPCAT \citep{Taylor:2005}.

\section*{Data Availability}

The LoTSS DR1 data used in this paper is publicly available, and can be found at \url{https://www.lofar-surveys.org/releases.html}. The cluster catalogue by \citet{WenHan2015} is also publicly available and is associated with the referenced paper. For access to the cluster-matched NAT data compiled from the catalogue, please contact KdV.




\bibliographystyle{mnras}
\bibliography{bibliography}








\bsp	
\label{lastpage}
\end{document}